# Effect of Sn-substitution on Thermoelectric Properties of Copper-based Sulfide, Famatinite $Cu_3SbS_4$

Yosuke Goto*, Yuki Sakai, Yoichi Kamihara, and Masanori Matoba

*Department of Applied Physics and Physico-Informatics, Faculty of Science and Technology, Keio University, Yokohama 223-8522, Japan*

Copper-based sulfide is an attractive material for Earth-abundant thermoelectrics. In this study, we demonstrate the effect of Sn-substitution on the electrical and thermal transport properties of fematinite $Cu_3SbS_4$ from 300 to 573 K. The carrier concentration is controlled in the range from $4 \times 10^{18}$ to $8 \times 10^{20}$ cm$^{-3}$ by Sn-substitution. The density-of-states effective mass is found to be ~3.0 $m_e$, assuming the single parabolic band model. The direct-type optical band gap is ~0.9 eV, which is consistent with the density functional theory calculation. The dimensionless figure of merit reaches 0.1 for Sn-doped samples at 573 K.

1.  Introduction

A thermoelectric generator/refrigerator is a solid-state device that can directly convert heat to electricity or pump heat using electricity without any gas or liquid working fluid.[1] The efficiency of these thermoelectric devices is primarily determined by Carnot efficiency and the material's dimensionless figure of merit $ZT = S^2 T \rho^{-1} \kappa^{-1}$, where $T$ is the absolute temperature, $S$ is the Seebeck coefficient, $\rho$ is the electrical resistivity, and $\kappa$ is the thermal conductivity.[2] Although many researchers have developed materials with $ZT > 1$,[3–5] further investigation toward high-performance thermoelectric materials is still required for more widespread use of thermoelectrics.[6] Recently, much attention has been paid to copper-based chalcogenides as promising $p$-type thermoelectric materials. Cubic $Cu_{2-x}Se$ exhibits low $\kappa$ because of its liquid-like phonon behavior.[7] Layered BiCuSeO, which also features reduced $\kappa$, originates from an alternating stack of a conducting CuSe layer and carrier-blocking BiO layer.[8] Such a feature is also observed in $LaCu_{1-\delta}SO$ and $LaOBiS_2$.[9,10] Furthermore, many compounds containing tetrahedrally coordinated copper-chalcogen units, such as $CuGaTe_2$,[11] $CuZnSnSe_4$,[12] and $Cu_3SbSe_4$,[13–15] were also reported as promising thermoelectric materials.

Among these compounds, $Cu_3SbSe_4$ could exhibit $ZT \sim 1.1$ via optimization of its chemical composition[16] despite its relatively high density-of-states effective mass (1.5 $m_e$ for Sn-doped system)[15], which usually leads to a lower thermoelectric quality factor.[17]

Several years ago, copper-based sulfides had rarely been investigated as thermoelectric materials, compared with tellurides and selenides, because of their poor $ZT$ value. However, recent studies on sulfide systems, such as $Cu_{2-x}S$,[18] $Cu_{12}Sb_4S_{13}$,[19,20] and $Cu_{26}V_2Ge_6S_{32}$,[21] have clearly shown that copper-based sulfides exhibit attractive thermoelectric properties. Compared with selenium, sulfur has the advantages of low cost, low toxicity, and high abundance (at least 1,000 times more abundant than selenium in the Earth's crust).[22]

In this study, the electrical and thermal transport properties of famatinite $Cu_3SbS_4$ with Sn-substitution were elucidated from 300 to 573 K. The crystallographic structure of $Cu_3SbS_4$ could be derived from zincblende through symmetry decay, as shown in Fig. 1.[23,24] Although the electrical transport of $Cu_3SbS_4$ with Ge-substitution was investigated up to 573 K,[25] thermal transport was limited to 300 K.[25,26] We demonstrate that the charge carrier transport of Sn-doped $Cu_3SbS_4$ is explained by simple electron counting and the single parabolic band (SPB) model, and lattice thermal conductivity is dominated by phonon–phonon Umklapp scattering above 300 K.

## 2. Experiments

Polycrystalline $Cu_3Sn_xSb_{1-x}S_4$ ($x$ = 0, 0.05, 0.10, and 0.15) was synthesized by a solid-state reaction using CuS, $Cu_2S$, $Sb_2S_3$, and $SnS_2$ as starting materials. Hereafter, the chemical composition of the samples was denoted by the nominal stoichiometry of these starting materials. First, the binary sulfides were prepared by heating a stoichiometric mixture of Cu (Kojundo Chemical, 99.9%), S (Sigma-Aldrich, 99.98%), Sb (Kojundo Chemical, 99.999%), and Sn (Kojundo Chemical, 99.99%) at 700 °C for 10 h in a sealed silica tube. Then, a stoichiometric mixture of the starting materials was pressed into pellets and heated at 450 °C for 40 h. After heat treatment, the sample was quenched in iced water. Finally, the sample was consolidated at 1 GPa and 400 °C by a cubic anvil press apparatus (Try-Engineering) using a pyrophyllite cell. The relative density of the obtained samples was calculated at more than 93%.

Powder X-ray diffraction (XRD) was performed using an X-ray spectrometer with a graphite monochromator (Rigaku, RINT 2500). The diffraction intensity was collected with CuKα radiation ($\lambda$ = 0.154060 nm) over a $2\theta$ range from 10 to 130° at a step width of 0.01°. Reference-grade Si powder (NIST SRM 640d) was utilized as an external reference. Each diffraction peak was fitted using a pseudo-Voigt function.[27] Lattice constants and corresponding statistical errors were calculated by a least-squares fitting method following Cohen[28] using more than 10 diffraction angles. Rietveld analysis was performed using the RIETAN-FP code.[29] The surface of the samples was examined using a scanning electron microscope (SEM; FEI Inspect S50).

The Hall coefficient ($R_H$) at room temperature was measured using the five-probe geometry under magnetic fields ($H$) up to $\mu_0 H = \pm 1$ Tesla. The Hall carrier concentration ($n_H$) was calculated as $n_H = 1/R_H e$, where $e$ is the charge of an electron in SI units ($e = 1.6021766 \times 10^{-19}$ C). The measurements of $\rho$, $S$, and $\kappa$ were conducted at temperatures up to 573 K using a lamp heating unit (Ulvac, MILA-5000). The value of $\rho$ was measured by the dc four-probe technique using Pt wires attached by silver paste (DuPont 6838) under a nitrogen atmosphere. $S$ was obtained from the slopes of the plots of the Seebeck voltages vs. temperature differences ($\Delta T$) measured with Pt–Pt/Rh 13% thermocouples. The value of $\kappa$ was obtained from the slopes of the plots of heat flux density vs. $\Delta T$, where $\Delta T$ was controlled using a strain gauge as a small heater. The $\kappa$ measurements were conducted under pressures of less than $10^{-3}$ Pa. The heat loss by radiation through the sample[30] was subtracted under the assumption that emissivity is independent of temperature and wavelength during the measurements of $\kappa$. The emissivity of 0.8 was employed on the basis of reflectivity ($R$) measurements, which were performed at room temperature using a spectrometer equipped with an integrating sphere (Hitachi High-tech, U-4100). Absorption spectra ($\alpha$) were converted from $R$ spectra using the Kubelka–Munk equation, $(1 - R)^2/2R = \alpha/s$, where $s$ denotes the scattering factor.[31]

The electronic structure calculation was performed using the plane-wave projector augmented-wave (PAW)[32,33] method implemented in the Vienna *ab initio* Simulation Package (VASP) code.[34,35] The exchange-correlation potential was approximated by the hybrid functionals using the Heyd, Scuseria, and Ernzerhof (HSE06) method.[36] The Brillouin zone was sampled by a $6 \times 6 \times 3$ Monkhorst–Pack grid,[37] and a cutoff of 450 eV was chosen for the plane-wave basis set. Hellmann–Feynman forces were reduced to 0.5 eV·nm$^{-1}$.

## 3. Results and Discussion

Figure 2(a) shows the XRD patterns of $Cu_3Sn_xSb_{1-x}S_4$. Almost all of the diffraction peaks could be assigned to those of the tetragonal phase, indicating that $Cu_3SbS_4$ is a dominant phase in the samples. However, diffraction due to the $SnO_2$ impurity is observed for the Sn-doped samples. The amount of $SnO_2$ increases with $x$ and it reaches 7.3 wt.% for $x = 0.15$ on the basis of Rietveld refinement. We note that partial cation disorder between Cu and Sb/Sn is not required for Rietveld refinement in contrast to $Cu_{3-\delta}SnS_4$,[38] where fitting results were shown in supporting information.[39] SEM images of polished samples indicate that the samples contain voids of ~1 μm diameter, as shown in Fig. 2(b).

The lattice parameters of $x = 0$ are calculated at $a = 0.53840(3)$ nm and $c = 1.0762(2)$ nm, as listed in Table I. The chemical composition dependences of $a$ and $c$ are less than the experimental error for $x \leq 0.1$, while that for $x = 0.15$ is distinctly smaller than that for the undoped sample. Such a nonlinear trend of lattice parameters was also observed in related compounds.[15,40–45]

The optical band gap was evaluated from absorption spectra converted from total diffuse reflectance spectra. Direct-type absorption edges were examined on the basis of the $(\alpha h\nu/s)^2$–$h\nu$ plot. As shown in Fig. 3, the band edge structure is observed at 0.9 eV, which is consistent with the electronic structure calculation, as described below.

Table I also summarizes the charge carrier transport at 300 K. The polarity of the Hall measurements confirms that the dominant carrier in the samples is a conducting hole. Figure 4(a) shows $n_H$ at 300 K versus $x$. An increase in $n_H$ with $x$ is in reasonable agreement with the expected value, assuming that each Sn generates one free hole, although the experimental $n_H$ is slightly lower than the expected value. Here, we used a fixed value of the Hall factor ($r_H$), $r_H = 1$.[46] Figure 4(b) shows $\rho$ as a function of $T$. The value of $\rho$ at $x = 0$ decreases with $T$ below 400 K, and the activation energy is calculated at 20 meV. An upturn in the $\rho$–$T$ plots above 400 K indicates degenerate conduction at these temperatures. The value of $\rho$ decreases to 2.7 mΩcm at 300 K as a result of the increased $n_H$ by Sn-substitution.

Figure 5 shows the $S$–$T$ plot. The value of $S$ at $x = 0$ slightly decreases with $T$ while that of the Sn-doped samples increases almost linearly up to 573 K, which is consistent with the theory of metal or heavily doped semiconductors.[47] Fig. 5(b) shows $S$ versus $n_H$ at 300 K, together with the theoretical curve generated using the SPB model with an effective mass ($m^*$) of 3.0 $m_e$, where $m_e$ is the rest mass of a free electron. Solutions to the Boltzmann transport equation within the relaxation time approximation are described by Eqs. (2) and (3) using the Boltzmann constant ($k_B = 8.61733 \times 10^{-5}$ eVK$^{-1}$), scattering parameter ($\lambda$), reduced chemical potential ($\eta^* = \eta/k_B T$), Fermi integral [$F(\eta^*)$], effective mass ($m^*$), and Planck constant ($h = 4.1356675 \times 10^{-15}$ eVs).[48,49] The value of $\lambda$ relates to the energy dependence of the carrier relaxation time, with $\tau = \tau_0 x^{\lambda-1/2}$, where $x$ is the reduced carrier energy ($x = E/k_B T$). In this article, we assume that the acoustic phonon scattering is dominant for charge carrier transport, $\lambda = 0$. The $i$th order of the Fermi integral is given by Eq. (4).

$$S = \frac{k_B}{e}\left(\frac{(2+\lambda)F_{1+\lambda}(\eta^*)}{(1+\lambda)F_\lambda(\eta^*)} - \eta^*\right) \quad (2)$$

$$n = 4\pi\left(\frac{2m^* k_B T}{h^2}\right)^{2/3} F_{1/2}(\eta^*) \quad (3)$$

$$F_i(\eta^*) = \int_0^\infty \frac{x^i}{1+\exp(x-\eta^*)}dx \quad (4)$$

Although the SnO$_2$ impurity is known to be an $n$-type conductor[50] and it might compensate the $p$-type conductivity of Cu$_3$SbS$_4$, the experimental results follow the same theoretical curve as the undoped (SnO$_2$-free) sample, indicating that the transport properties of the samples are dominated by the Cu$_3$SbS$_4$ phase.

Figure 6(a) shows the temperature dependence of the total thermal conductivity $\kappa$. The lattice thermal conductivity ($\kappa_l$) was calculated by subtracting the electronic thermal conductivity ($\kappa_e$) from $\kappa$. The value of $\kappa_e$ is described by the Wiedemann–Franz law, $\kappa_e = L\rho^{-1}T$, where $L$ is the Lorenz number obtained by Eq. (5) within the SPB model. The value of $\eta^*$ was calculated using experimental Seebeck coefficients and Eq. (2).

$$L = \left(\frac{k_B}{e}\right)^2 \frac{(1+\lambda)(3+\lambda)F_\lambda(\eta^*)F_{\lambda+2}(\eta^*) - (2+\lambda)^2 F_{\lambda+1}(\eta^*)^2}{(1+\lambda)^2 F_\lambda(\eta^*)^2} \quad (5)$$

The value of $\kappa_l$ decreases with $T$ following $T^{-1}$ behavior, indicating that the phonon–phonon Umklapp scattering is dominant in the thermal transport at these temperatures.[51] The value of $\kappa_l$ at $x = 0.15$ at 300 K was distinctly lower than that of the other samples, which might be attributed to point-defect scattering for the solid solution. Figure 7 shows the dimensionless figure of merit versus temperature. The $ZT$ of the samples increases with $T$. The $ZT$ at $x = 0$ was 0.03 at 573 K, and it increased to 0.1 for the Sn-doped samples. This value is comparable to that of $Cu_2ZnSnS_4$ at these temperatures.[52]

Figure 8(a) shows the calculated total and partial density of states for $Cu_3SbS_4$. The band gap, which is determined by the energy difference between the valence band maximum (VBM) and the conduction band minimum (CBM), is calculated at 1.0 eV, which is consistent with the experimental results (Fig. 3). The VBM is composed of a hybridization between the Cu $3d$ orbitals and the S $3p$ orbitals, whereas the CBM is composed of the Sb $5s$ and S $3p$ orbitals. The band dispersion is shown in Fig. 8(b). The VBM at the $\Gamma$ point consists of three bands with different effective masses. Parabolic fits were used to estimate the $m^*$ of each band. The $m^*$ ranges from 0.4 to 4.2 $m_e$, which is consistent with the experimentally obtained value using the SPB model, 3.0 $m_e$. The experimental value using the SPB model is attributed to the averaged value contributed by these three degenerate bands with different effective mass. The calculated density of states and band dispersion are similar to those previously reported, except for the value of the band gap.[53,54] The band gap values of ~0.6 and ~1.0 eV were reported using the PBE functional[53] and PBE + $U$ functional ($U_{eff}$ = 15 eV),[54] respectively. It is almost a consensus that the PBE functional underestimates the band gap, whereas the HSE06 hybrid functional, which is used in this study, provides a band gap with reasonable accuracy in related compounds.[55]

## 5. Conclusions

Polycrystalline $Cu_3SbS_4$ with Sn-substitution was prepared to examine the electrical and thermal transport properties. The Hall carrier concentration was controlled in the range from $4 \times 10^{18}$ to $8 \times 10^{20}$ cm$^{-3}$ by Sn-substitution. The density-of-states effective mass was found to be 3.0 $m_e$ using the SPB model. The phonon–phonon Umklapp scattering was dominant for thermal transport at 300–573 K. The dimensionless figure of merit was 0.1 for the Sn-doped samples. The direct-type band gap was ~0.9 eV, which was determined from both theory and experiments.

**Acknowledgment**

We thank Dr. Ikuya Yamada of Osaka Prefecture University for the use of the high-pressure apparatus. This work was partially supported by research grants from Keio University, the Keio Leading-edge Laboratory of Science and Technology (KLL), Japan Society for Promotion of Science (JSPS) KAKENHI Grant Number 26400337, and Hitachi Metals Materials Science Foundation (HMMSF).*E-mail: ygoto@keio.jp

Table caption

Table I. Lattice parameters ($a$ and $c$), Hall carrier concentration ($n_H$), electrical resistivity ($\rho$), Hall carrier mobility ($\mu_H$), and Seebeck coefficient ($S$) of $Cu_3Sn_xSb_{1-x}S_4$ at 300 K. The values in parentheses are the statistical errors. Other errors such as temperature fluctuations (< 1 K) should be considered for $a$ and $c$.

Figure captions

Fig. 1. (Color online) Crystallographic structure of famatinite $Cu_3SbS_4$ belonging to the tetragonal $I\bar{4}2m$ space group.

Fig. 2. (Color online) (a) X-ray diffraction (XRD) patterns of $Cu_3Sn_xSb_{1-x}S_4$. Vertical bars at the bottom represent the calculated Bragg positions of $Cu_3SbS_4$. Arrows denote the diffraction due to $SnO_2$ as a secondary phase. (b) Scanning electron microscopy image of the polished surface for $x = 0$.

Fig. 3. (Color online) Absorption spectra of $Cu_3SbS_4$ converted from reflectivity spectra. The direct-type absorption edge was evaluated from the onset of the $(\alpha h\nu/s)^2 - h\nu$ plot.

Fig. 4. (Color online) (a) Hall carrier concentration ($n_H$) of $Cu_3Sn_xSb_{1-x}S_4$ at room temperature. The dashed line represents the expected value assuming each Sn provides one free hole. (b) Electrical resistivity ($\rho$) as a function of temperature ($T$).

Fig. 5. (Color online) (a) Seebeck coefficient ($S$) as a function of temperature ($T$) for $Cu_3Sn_xSb_{1-x}S_4$. (b) $S$ versus Hall carrier concentration ($n_H$) at 300 K. The dashed line was generated using a single parabolic band model with an effective mass of 3.0 $m_e$.

Fig. 6. (Color online) Temperature dependences of (a) total thermal conductivity ($\kappa$) and (b) lattice thermal conductivity ($\kappa_l$). The dashed line represents the $\kappa_l \propto T^{-1}$ behavior corresponding to Umklapp phonon–phonon scattering.

Fig. 7. (Color online) Dimensionless figure of merit ($ZT$) versus temperature ($T$) for $Cu_3Sn_xSb_{1-x}S_4$.

Fig. 8. (Color online) (a) Total and partial density of states (DOS) of $Cu_3SbS_4$. Energy is aligned by the Fermi level at 0 eV. (b) Band structure of $Cu_3SbS_4$. The effective mass of the valence band indexed as A–E is calculated at 4.2, 1.1, 0.4, 0.7, and 0.4 $m_e$, respectively, using parabolic fit.

Table I

| $x$ | $a$ (nm) | $c$ (nm) | $n_H$ ($10^{20}$ cm$^{-3}$) | $\rho$ (mΩcm) | $\mu_H$ (cm$^2$V$^{-1}$s$^{-1}$) | $S$ (μVK$^{-1}$) |
|---|---|---|---|---|---|---|
| 0 | 0.53840(3) | 1.0762(2) | 0.042(1) | 249.0 | 6.0 | 564 |
| 0.05 | 0.53836(4) | 1.0763(2) | 1.04(5) | 29.1 | 2.1 | 202 |
| 0.10 | 0.53838(3) | 1.0762(1) | 6.0(7) | 6.2 | 1.7 | 82 |
| 0.15 | 0.53823(3) | 1.0759(1) | 7.8(5) | 2.7 | 2.9 | 71 |

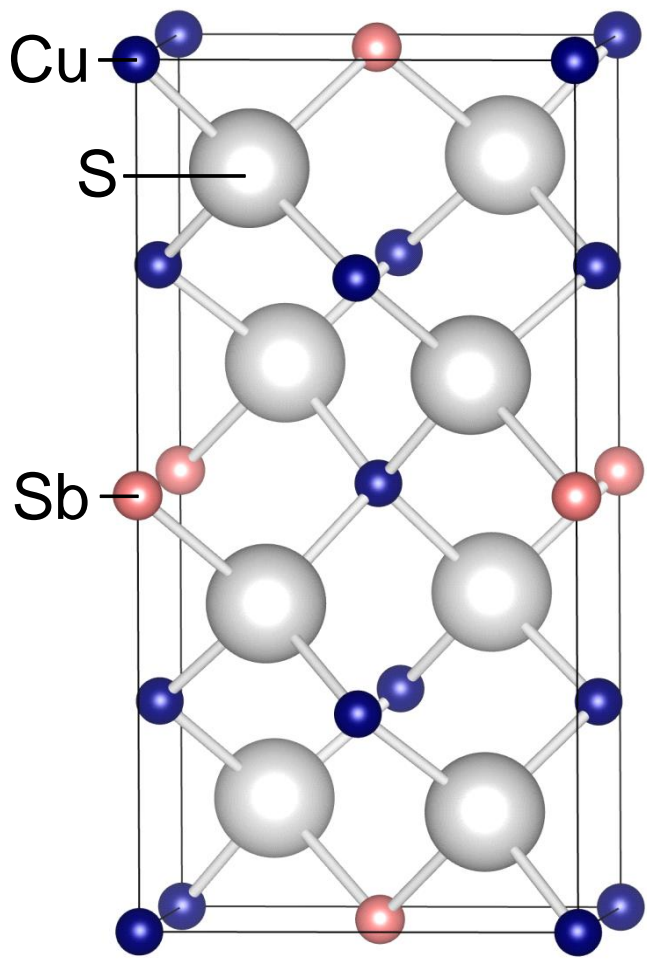

Fig. 1

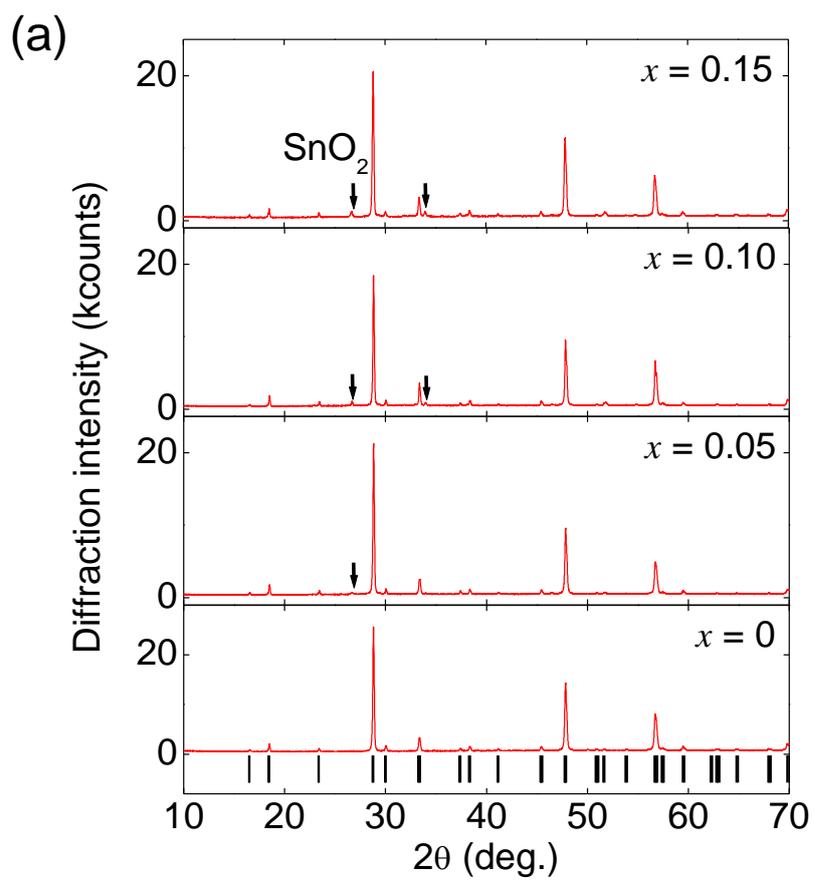

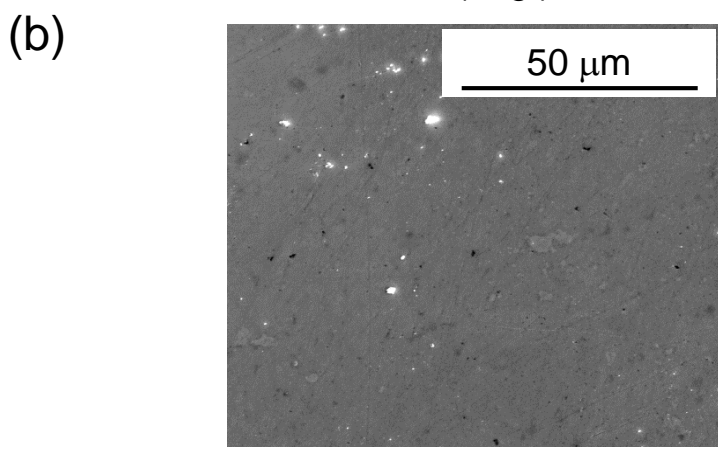

Fig. 2

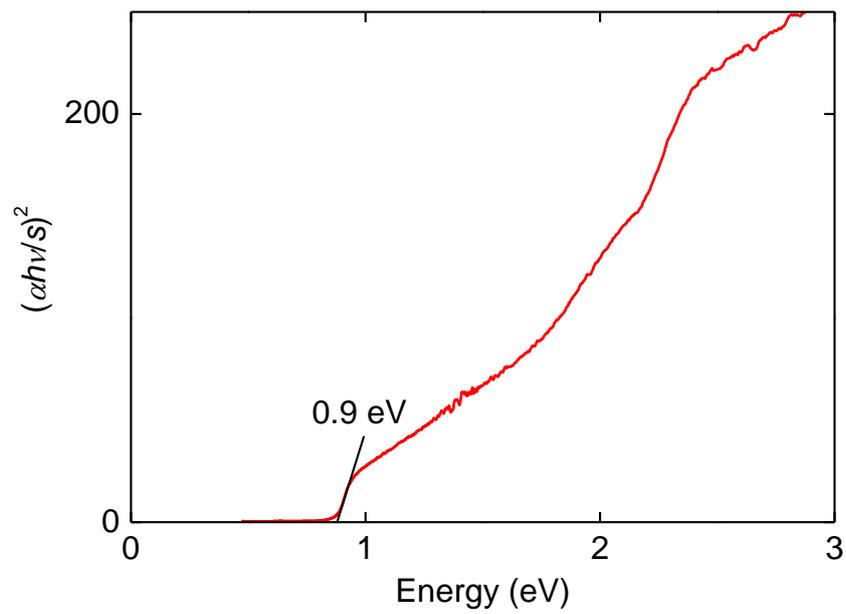

Fig. 3

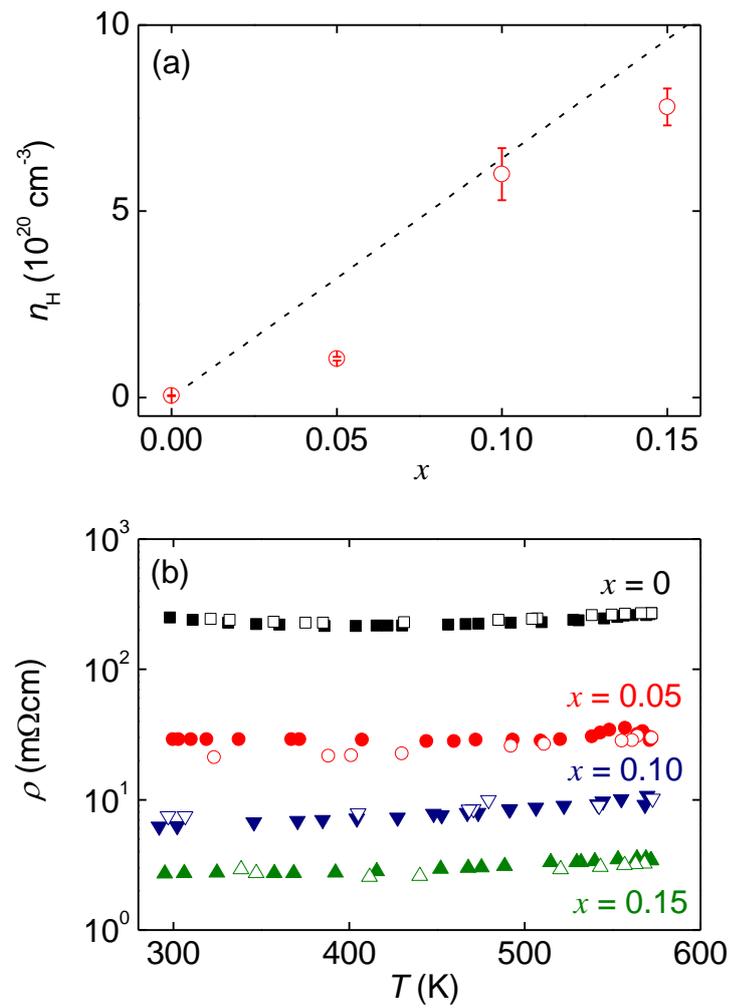

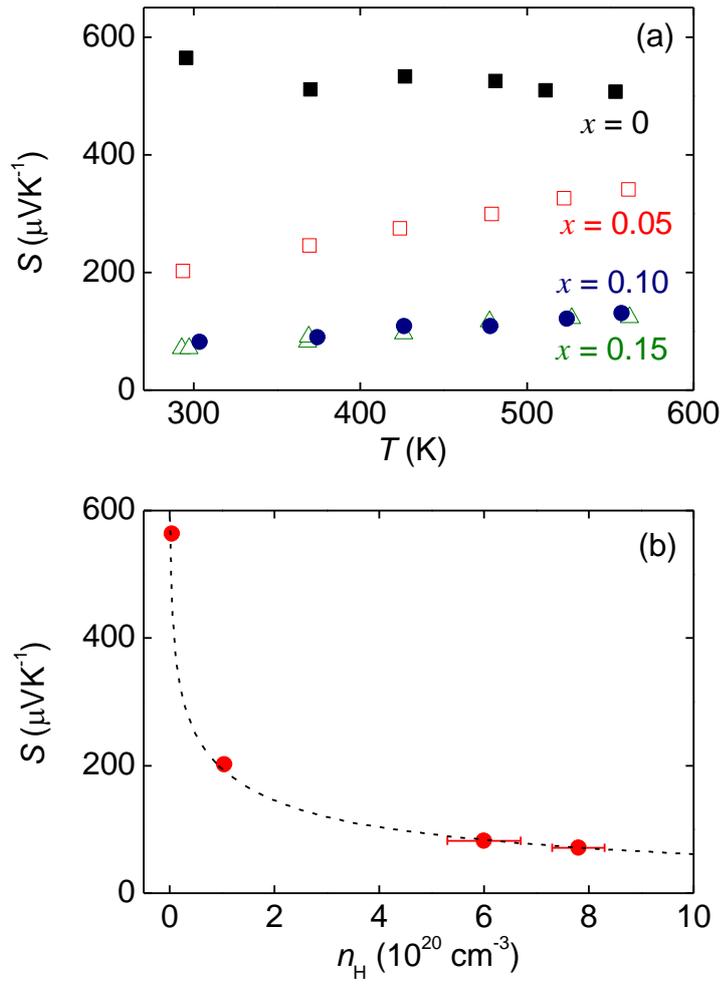

Fig. 5

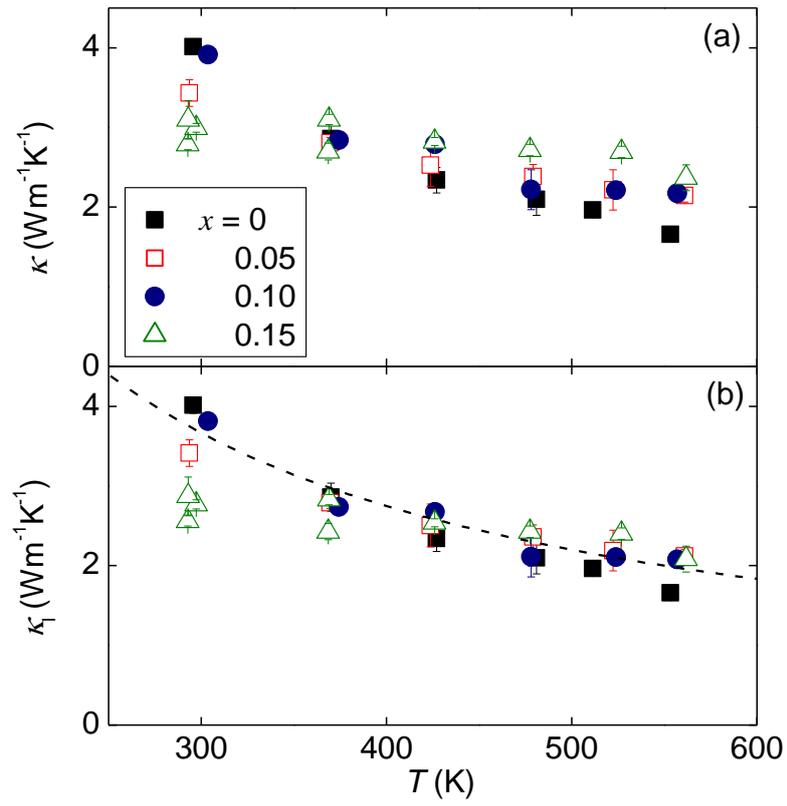

Fig. 6

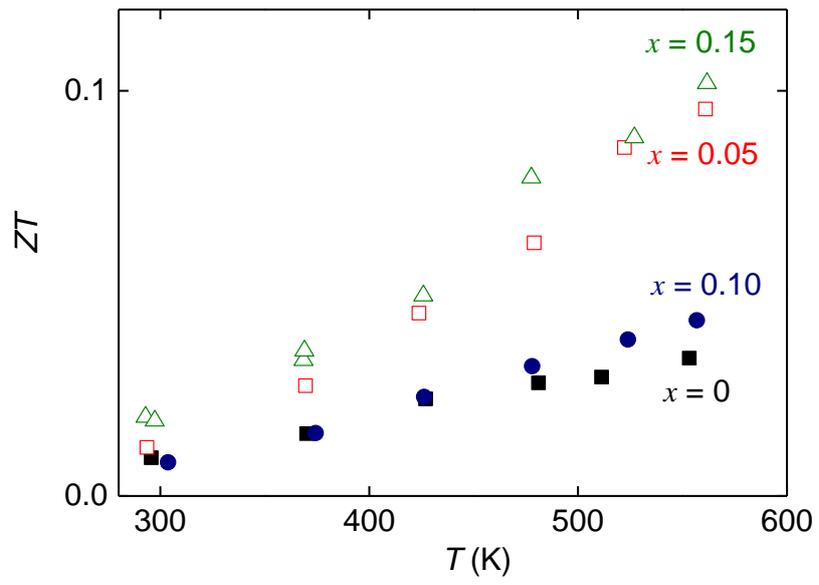

Fig. 7

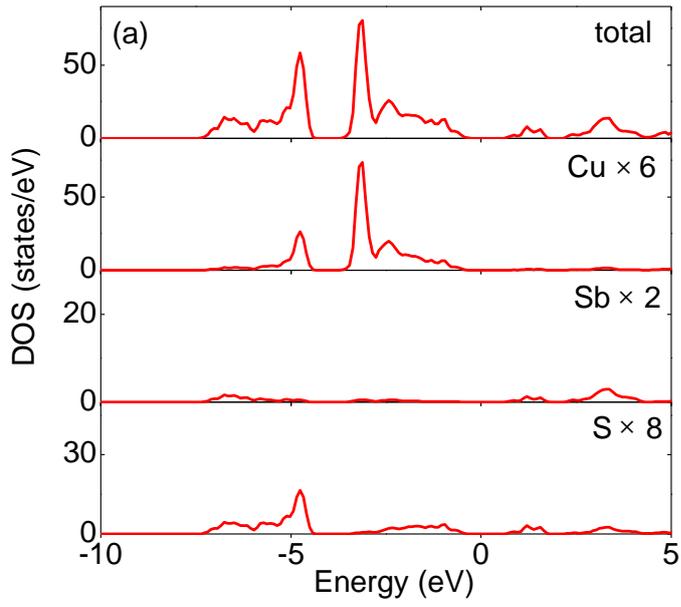

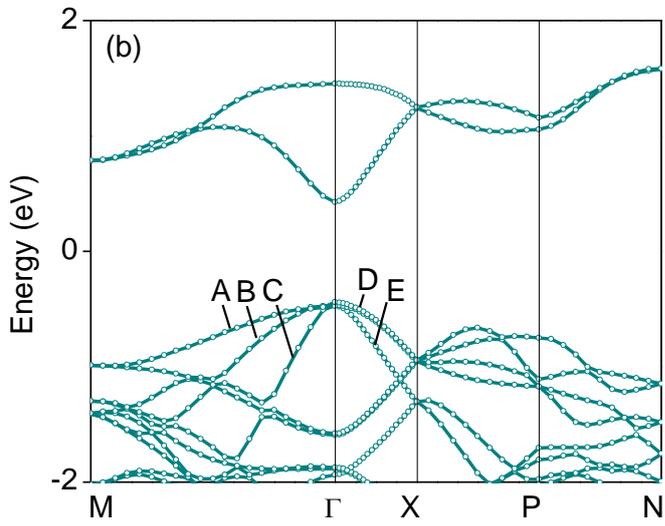

Fig. 8